\documentclass[conference]{IEEEtran}
\IEEEoverridecommandlockouts
\usepackage{amsthm}
\theoremstyle{definition}
\newtheorem{definition}{Definition}
\usepackage{cite}
\usepackage{amsmath,amssymb,amsfonts}
\usepackage{algorithmic}
\usepackage{graphicx}
\usepackage{textcomp}
\usepackage{xcolor}
\usepackage[colorlinks=true, linkcolor=blue, citecolor=blue]{hyperref}
    
\def\BibTeX{{\rm B\kern-.05em{\sc i\kern-.025em b}\kern-.08em
    T\kern-.1667em\lower.7ex\hbox{E}\kern-.125emX}}

\begin{document}

\title{Low Probability of Target Detection in Wireless Sensing by Harnessing Reconfigurable-antenna based Reflectors
}


\author{
\IEEEauthorblockN{
Yifan Zhang\IEEEauthorrefmark{1},
Zhentian Zhang\IEEEauthorrefmark{2},
Jäntti Riku\IEEEauthorrefmark{1},
Yichuan Wang\IEEEauthorrefmark{3}
Xingwang Li\IEEEauthorrefmark{3}
Tuo Wu\IEEEauthorrefmark{4}
}
\IEEEauthorblockA{\IEEEauthorrefmark{1}Department of Information and Communications Engineering, Aalto University, Espoo, Finland}

\IEEEauthorblockA{\IEEEauthorrefmark{2}School of Cyber Engineering, Hong Kong Polytechnic University, Hong Kong SAR, China}
\IEEEauthorblockA{\IEEEauthorrefmark{3}College
of Computer Science and Engineering, Xi’an University of Technology, Xi'an, China}
\IEEEauthorblockA{\IEEEauthorrefmark{3}School of Physics and Electronic Information Engineering, Henan Polytechnic University, Jiaozuo, China}
\IEEEauthorblockA{\IEEEauthorrefmark{4}School of Electronic and Information Engineering, South China University of Technology, Guangzhou, China}

\IEEEauthorblockA{Email: \{yifan.1.zhang, riku.jantti\}@aalto.fi; zhentianzhangzzt@gmail.com;
 chuan@xaut.edu.cn; lixingwang@hpu.edu.cn;\\ wutuo@scut.edu.cn
}
} 

\maketitle

\begin{abstract}
In this paper, we introduce low probability of target detection (LPTD) as a new security objective for wireless sensing systems, where the existence of a target is concealed from passive wardens. Unlike conventional communication security and covert communication, LPTD aims to protect the sensed target itself from malicious detection, rather than concealing transmitted signals or communication activities. Specifically, we propose a reconfigurable-antenna-based reflector-assisted target covertness scheme, where passive reflectors deployed around the target are harnessed as artificial scattering sources to confuse the warden. By adjusting their reflection coefficients, these reflectors reshape the local scattering environment and mask the target-induced echo. We then analyze the warden's optimal detection strategy and derive the false-alarm and missed-detection probabilities, followed by a closed-form expression for its minimum total detection error probability. Numerical results validate the analysis and show that sufficient cooperation among reflectors can increase the warden's minimum total detection error probability to approximately 0.94--0.96, making target detection close to random guessing.
\end{abstract}

\begin{IEEEkeywords}
Low probability of target detection, wireless sensing security, reconfigurable-antenna based reflectors, covert target, and detection error probability.
\end{IEEEkeywords}

\section{Introduction}
Wireless sensing technologies have provided ubiquitous target awareness, but it also exposes physical targets to unauthorized observation. By reusing radio signals for detection, localization, and environmental perception, wireless sensing has become a promising technology for next-generation networks \cite{zhang2025backscatter}. However, the target's privacy in wireless sensing is inherently exposed to the open propagation environment. Since sensing waveforms and target echoes naturally radiate through space, a passive warden can detect the target's presence and further estimate its sensitive sensing information by directly sensing the target's physical response \cite{10587082}, leading to serious privacy leakage. Therefore, preventing the illegal detection of targets becomes a fundamental security challenge in wireless sensing systems.

Existing wireless security techniques cannot directly address this target detection threat, as they primarily focus on securing transmitted data or hiding communication behavior. Specifically, traditional physical-layer security methods, such as secure beamforming \cite{10436390}, artificial noise \cite{zhang2025artificial}, and cooperative jamming \cite{10109228}, usually require the passive warden's CSI or location for directional jamming design, and suffer from scalability issues due to extra power consumption or dedicated deployment \cite{11122381}. Meanwhile, covert communications, also known as low probability of detection (LPD) communications, have been widely studied to hide confidential message transmission from a warden through techniques such as low-power transmission \cite{10906391}, spread-spectrum signaling \cite{11318877}, artificial noise, and backscatter-assisted reflection control \cite{zhang2024covert}. However, since the warden detects the target by capturing its natural signal reflections, the target remains exposed regardless of how covert the transmitted message is. In summary, existing solutions are still insufficient to hinder the detection of targets in wireless sensing networks.



To protect the target from being detected, this work proposes a novel scheme that leverages reconfigurable-antenna-based reflectors to achieve low probability of target detection (LPTD) against wardens. Specifically, the proposed scheme harnesses widely distributed reflectors around the target, which can be implemented in various forms, such as ambient backscatter devices \cite{11667367}, reconfigurable surfaces \cite{9696209}, and reconfigurable corner reflectors \cite{10955353}, to act as artificial scattering sources to confuse the warden. By appropriately configuring their reflection coefficients or electromagnetic responses, these reflectors reshape the local scattering environment \cite{Zhang2026AmbShield,11071870}, thereby reducing the statistical distinguishability between the target-induced scattering response and the surrounding reflected signals in the warden's observation. In this way, the target echo is concealed within the uncertainty introduced by the surrounding reflections, making it difficult for a passive warden to reliably determine the presence of the true target. Specifically, the main contributions of this paper are summarized as follows.

\begin{figure}[tp]
    \centering
    \includegraphics[width=0.9\linewidth]{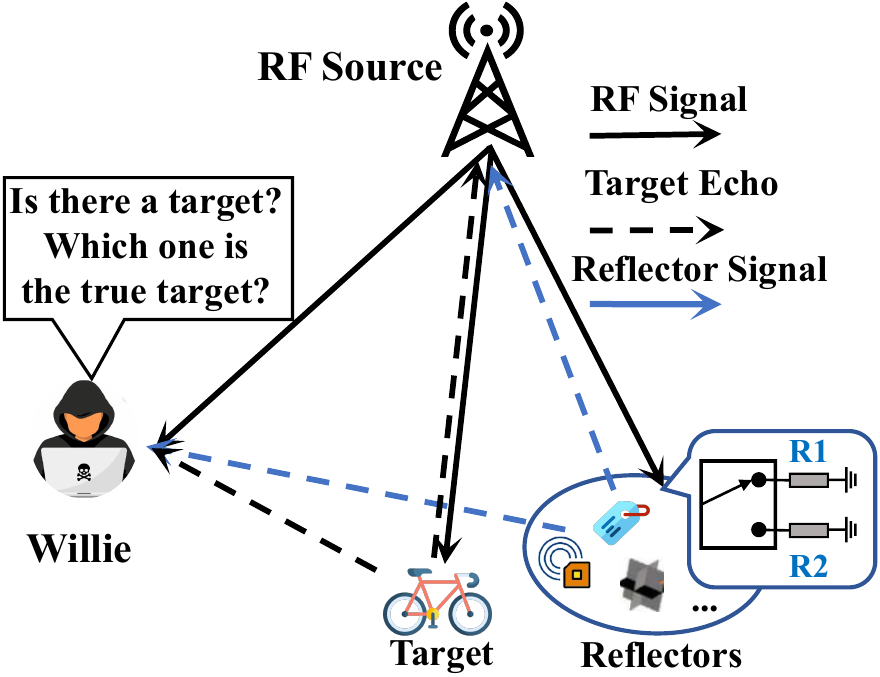}
    \caption{The system model, where a malicious target detection warden is confused by the signals from reflectors in sensing the existence of a true target.}
    \label{syse}
   \vspace{-0.5cm}
\end{figure}

 \begin{itemize} 
 
 \item We propose an reflector‑assisted target covertness scheme to achieve LPTD against a target-detection warden. Specifically, we model a system where a multi‑antenna monostatic radar transceiver illuminates the target area, while a passive warden (Willie) attempts to detect the target’s presence. The reflectors act as controllable reflectors to obfuscate Willie’s observation. Within this model, we extend the concept of LPD and formally define LPTD based on Willie’s minimum total detection error probability, which, to the best of our knowledge, is the first work in IoT sensing environments.
 
 \item We formulate the target detection problem from the warden's perspective and derive a closed-form expression for the minimum total detection error probability. By modeling a binary hypothesis-testing problem of target detection, we characterize Willie’s optimal target detection strategy, derive the false-alarm and missed-detection probabilities, and obtain a closed-form detection error boundary.

  \item  We verify the effectiveness of the proposed scheme through both theoretical derivation and simulation.
  The results demonstrate the effect of key system parameters on the LPTD performance, including the number of reflectors, transmit SNR, and scattering coherence. It is evident that sufficient reflector cooperation can increase the warden's minimum total detection error probability to about 0.94--0.96, rendering Willie's target detection nearly random guessing.
 
\end{itemize}

\textit{Notations:} Boldface lower-case letters denote vectors. 
The operators $(\cdot)^T$ and $(\cdot)^H$ represent the transpose and the Hermitian transpose, respectively. 
$|\cdot|$ denotes the absolute value of a complex scalar.
$\mathbb{E}\{\cdot\}$, $\Pr(\cdot)$, and $\Pr(\cdot|\cdot)$ represent the statistical expectation, probability, and conditional probability operators, respectively.
$x \sim \mathcal{CN}(\mu, \sigma^2)$ indicates that the random variable $x$ follows a circularly symmetric complex Gaussian (CSCG) distribution with mean $\mu$ and variance $\sigma^2$.

\section{Problem Formulation}

\subsection{System Model}

As shown in Fig.~\ref{syse}, we consider a wireless sensing system consisting of a legitimate base station (BS), one target, a target-detection warden named Willie, and $K$ reflectors deployed around the target. The BS transmits sensing waveforms to illuminate the target area, while Willie passively observes the wireless environment to determine whether the target is present. The reflectors are low-power passive devices that manipulate incident radio-frequency (RF) signals through controllable load impedances \cite{luoniya}.

The BS is equipped with $M$ transmit antennas, while Willie and each reflector have a single antenna \cite{11365966}. The reflector set is denoted by $\mathcal{K}=\{1,\ldots,K\}$. A sensing frame contains $L$ symbols \cite{11155198}, where $L$ is sufficiently large such that noise fluctuations and signal-noise cross terms vanish. The BS transmit signal at the $\ell$-th symbol is
\begin{equation}
    \mathbf{x}[\ell]=\mathbf{w}s[\ell], \quad \ell=1,\ldots,L,
    \label{eq:bs_signal}
\end{equation}
where $\mathbf{w}\in\mathbb{C}^{M\times1}$ is the beamforming vector satisfying $\|\mathbf{w}\|^2\leq P_{\max}$, and $s[\ell]$ is the normalized sensing symbol with $\frac{1}{L}\sum_{\ell=1}^{L}|s[\ell]|^2=1$.

All channels follow a block-fading model, remaining constant within each sensing frame and varying independently across frames. Let $\mathbf{h}_{Bi}\in\mathbb{C}^{M\times1}$ denote the BS-to-node channel for $i\in\{T,W,1,\ldots,K\}$, where $T$, $W$, and $k\in\mathcal{K}$ denote the target, Willie, and the $k$-th reflector, respectively. Let $h_{ij}\in\mathbb{C}$ denote the scalar channel between nodes $i$ and $j$. Specifically, $\mathbf{h}_{Bi}=\sqrt{\ell(d_{Bi})}\tilde{\mathbf{h}}_{Bi}$ and $h_{ij}=\sqrt{\ell(d_{ij})}\tilde{h}_{ij}$, where $\tilde{\mathbf{h}}_{Bi}\sim\mathcal{CN}(\mathbf{0},\mathbf{I}_M)$ and $\tilde{h}_{ij}\sim\mathcal{CN}(0,1)$ are independent small-scale fading components. The large-scale path loss is modeled as $\ell(d)=C_0(d/d_0)^{-\alpha}$, where $C_0$ is the reference path loss at distance $d_0$, and $\alpha$ is the path-loss exponent.

\subsection{Reflection Signal of Reflectors}

The harnessed reconfigurable-antenna-based reflectors are deployed around the target to manipulate and reflect the incident RF signal. These reflectors can be implemented by ambient backscatter devices, RIS elements, or reconfigurable corner reflectors. Unlike active jammers, they do not generate new RF signals on their own. Instead, they reshape the surrounding scattering environment by controlling their electromagnetic reflection responses \cite{famaey2026survey}.

For a reconfigurable antenna, the reflected signal can be controlled by adjusting the load impedance connected to the antenna. Specifically, the antenna reflection coefficient is given by
\begin{equation}
    \Gamma_k[\ell]
    =
    \frac{R_{L,k}[\ell]-R_{A,k}^{*}}
    {R_{L,k}[\ell]+R_{A,k}},
    \label{eq:antenna_reflection_coefficient}
\end{equation}
where $R_{L,k}[\ell]$ denotes the configurable load impedance of the $k$-th reflector at the $\ell$-th sensing symbol, $R_{A,k}$ denotes its antenna impedance. By switching or continuously tuning $R_{L,k}[\ell]$, the reflector can accordingly control the amplitude and phase of the reflected signal. Such impedance-based control provides a unified physical interpretation for different passive or nearly passive reflecting implementations.

Accordingly, we model the effective complex reflection coefficient of the $k$-th reflector at the $\ell$-th sensing symbol as
\begin{equation}
    \beta_k[\ell]
    =
    \rho_k[\ell]e^{j\phi_k[\ell]},
    ,
    \label{eq:reflection_coefficient}
\end{equation}
where $\rho_k[\ell]$ and $\phi_k[\ell]$ denote the effective reflection amplitude and phase shift, respectively 
    $0\leq \rho_k[\ell]\leq \rho_{k,\max},
    \phi_k[\ell]\in[0,2\pi)$. The coefficient $\beta_k[\ell]$ captures the overall configurable reflection response, including the effect of the antenna-load impedance and other implementation-dependent losses. Typically, the reflection coefficients remain fixed within a sensing frame, i.e., $\beta_k[\ell]=\beta_k$, while they can be reconfigured across different frames to alter the local scattering environment.

The signal reflected by the $k$-th reflector is therefore expressed as
\begin{equation}
    x_k[\ell]
    =
    \beta_k[\ell]\mathbf{h}_{Bk}^{H}\mathbf{w}s[\ell],
    \label{eq:reflector_reflected_signal}
\end{equation}
where $\mathbf{h}_{Bk}^{H}\mathbf{w}s[\ell]$ represents the incident signal received by the $k$-th reflector from the BS.

For notational convenience, the reflection coefficients of all $K$ reflectors at the $\ell$-th sensing symbol are collected into
\begin{equation}
    \boldsymbol{\beta}[\ell]
    =
    [\beta_1[\ell],\ldots,\beta_K[\ell]]^T.
    \label{eq:reflection_vector}
\end{equation}

\subsection{Willie's Target Detection}

Willie's goal is to infer the target's existence from the received sensing signals. 
The general received signal at Willie during the $\ell$-th sensing symbol is given by
\begin{align}
    y_W[\ell]
    =
    \mathbf{h}_{BW}^{H}\mathbf{w}s[\ell]
    +
     \alpha_T h_{TW} \mathbf{h}_{BT}^{H}\mathbf{w}s[\ell] \nonumber \\ 
    +
    \sum_{k=1}^{K}
    h_{kW}x_k[\ell]
    +
    n_W[\ell],
    \label{eq:willie_received_signal_general}
\end{align}
where $\alpha_T \sim \mathcal{CN}(0, \sigma_t^2)$ represents the radar cross section (RCS) \cite{zhang2026ambsentry} of the sensing target, and $n_W[\ell]\sim\mathcal{CN}(0,\sigma_W^2)$ is the additive white Gaussian noise (AWGN) at Willie. Due to the severe multiplicative path loss, the multi-cascaded signal between the target and the reflectors is neglected in the following performance analysis.

To detect the target, Willie formulates a binary hypothesis testing problem \cite{10497104}. Define $\mathcal{H}_0$ as the target is absent, and $\mathcal{H}_1$ as the target is present. Therefore, Willie's received signal under $\mathcal{H}_0$ and $\mathcal{H}_1$ is expressed as 
\begin{equation}
    y_W[\ell] = 
    \begin{cases} 
    \begin{aligned}
       & \mathbf{h}_{BW}^{H}\mathbf{w}s[\ell] + \sum\limits_{k=1}^{K} h_{kW}\beta_k[\ell] \mathbf{h}_{Bk}^{H}\mathbf{w}s[\ell] \\&+ n_W[\ell],
        \end{aligned}
        & \mathcal{H}_0, \\ 
        \begin{aligned}
            &\mathbf{h}_{BW}^{H}\mathbf{w}s[\ell] + \alpha_T h_{TW} \mathbf{h}_{BT}^{H}\mathbf{w}s[\ell] \\
            &+ \sum\limits_{k=1}^{K} h_{kW}\beta_k[\ell] \mathbf{h}_{Bk}^{H}\mathbf{w}s[\ell] + n_W[\ell],
        \end{aligned} & \mathcal{H}_1. 
    \end{cases}
    \label{eq:willie_hypotheses}
\end{equation}

\subsection{Design Goals} 
The objective of our scheme design is to achieve LPTD at Willie,
characterizing whether the reflectors can conceal the existence of the target from Willie. The reflectors reshape the scattering environment around the target, making the received energy at Willie under $\mathcal{H}_0$ and $\mathcal{H}_1$ difficult to distinguish. Specifically, we define LPTD as

\begin{definition}
Let $P^*_E$ denote Willie's optimal minimum total detection error probability after optimizing his detection threshold. The target is said to achieve LPTD (or covert target) if 
\begin{equation} 
    P_E^* \geq 1-\epsilon, \label{eq:lptd_constraint} 
\end{equation} 
where $\epsilon \in [0,1]$ specifies the desired LPTD or target covertness level.
\end{definition}

\section{Performance Analysis for Willie's Detection Error Boundary}

In this section, we derive Willie's detection error boundary under the proposed reflector-assisted target covertness model.

\subsection{Willie'sDetection Strategy}

Since the BS sensing waveform is known, Willie can remove the deterministic direct BS-Willie component if the corresponding channel is available \cite{11202391}. The residual received signal is defined as
\begin{equation}
    \tilde{y}_W[\ell]
    =
    y_W[\ell]
    -
    \mathbf{h}_{BW}^{H}\mathbf{w}s[\ell].
    \label{eq:willie_residual_signal}
\end{equation}

Then, Willie performs a binary hypothesis test based on the residual observation vector
    $\tilde{\mathbf{y}}_W
    =
    [\tilde{y}_W[1],\ldots,\tilde{y}_W[L]]^T$. According to the Neyman--Pearson lemma \cite{poor2013introduction}, for a given false alarm constraint, Willie's optimal detector is the likelihood ratio test, given by
\begin{equation}
    \Lambda(\tilde{\mathbf{y}}_W)
    =
    \frac{
    f_{\tilde{\mathbf{y}}_W|\mathcal{H}_1}
    \left(
    \tilde{\mathbf{y}}_W|\mathcal{H}_1
    \right)
    }{
    f_{\tilde{\mathbf{y}}_W|\mathcal{H}_0}
    \left(
    \tilde{\mathbf{y}}_W|\mathcal{H}_0
    \right)
    }
    \mathop{\gtrless}_{\mathcal{H}_0}^{\mathcal{H}_1}
    \eta,
    \label{eq:willie_lrt}
\end{equation}
where $\eta$ is the likelihood ratio threshold.

Since Willie does not know the instantaneous reflector reflection coefficients and cascaded reflector channels \cite{Zhang2026AmbShield}, he cannot coherently separate the target-induced echo from the reflector-induced reflections. Therefore, we consider a non-coherent radiometer detector, which is widely adopted in detection problems with unknown instantaneous CSI and a single antenna Willie \cite{7964713}. Under this detector, Willie makes his decision based on the received residual energy, i.e., 
    $T_W
    =
    \frac{1}{L}
    \sum_{\ell=1}^{L}
    \left|
    \tilde{y}_W[\ell]
    \right|^2$. Therefore, the corresponding decision for the existence of the target is
\begin{equation}
    T_W
    \mathop{\gtrless}_{\mathcal{H}_0}^{\mathcal{H}_1}
    \tau,
    \label{eq:willie_energy_test}
\end{equation}
where $\tau$ is the energy detection threshold.

Let $\mathbf{w}=\sqrt{P_B}\mathbf{v}$, where $\mathbf{v}$ denotes the transmit beamforming vector satisfying $\|\mathbf{v}\|^2=1$. Define the equivalent reflector-induced scattering channel as
    $g_A
    =
    \sum_{k=1}^{K}
    \beta_k
    h_{kW}
    \mathbf{h}_{Bk}^{H}\mathbf{v}$, and the direct target-induced scattering channel is defined as
 $g_T
    =
    \alpha_T h_{TW}
    \mathbf{h}_{BT}^{H}\mathbf{v}$, Willie's residual signal can be written as
\begin{align}
    &\mathcal{H}_0:
    \quad
    \tilde{y}_W[\ell]
    =
    \sqrt{P_B}g_A s[\ell]
    +
    n_W[\ell],\\
    & \mathcal{H}_1:
    \quad
    \tilde{y}_W[\ell]
    =
    \sqrt{P_B}
    \left(
    g_A+g_T
    \right)
    s[\ell]
    +
    n_W[\ell].
    \label{eq:residual_h1_corrected}
\end{align}

By the law of large numbers for the transmitted samples, the radiometer output converges to
\begin{align} 
    &\mathcal{H}_0:
    \quad
    T_W
    =
    P_B|g_A|^2+\sigma_W^2,\\ \label{eq:TW_H0_corrected}
    &\mathcal{H}_1:
    \quad
    T_W
    =
    P_B|g_A+g_T|^2+\sigma_W^2.
\end{align}

\subsection{False Alarm and Missed Detection Probabilities}

 Although the channels themselves are independent, reflectors may exhibit spatial correlation or deliberate phase alignment, thereby resulting in a non-zero.
To account for possible correlation among the reflector-reflected paths, we introduce a partial coherence factor $\chi\in[0,1]$ among different reflector-reflected paths. A larger $\chi$ indicates stronger coherent combining among the reflector-reflected signals at Willie.  Furthermore, we approximate the aggregate reflector-induced scattering channel as an equivalent circularly symmetric complex Gaussian random variable. This approximation is motivated by the central limit theorem \cite{tse2005fundamentals}, since $g_A$ is formed by the superposition of multiple independent reflected paths with random amplitudes and phases. Thus, we have
\begin{equation}
    g_A
    \sim
    \mathcal{CN}
    \left(
    0,
    \Omega_A(K,\chi,M)
    \right),
\end{equation}
where $\Omega_A(K,\chi,M)$ denotes the normalized average reflector-induced scattering strength. To capture the partial coherence among different reflector-reflected paths, define
    $q_k(M)
    =
    \bar{\rho}_k^2
    G_k(M)
    \mathcal{L}(d_{Bk})
    \mathcal{L}(d_{kW})$,
where $\bar{\rho}_k^2=\mathbb{E}\{|\beta_k|^2\}$ is the average reflection power of the $k$-th reflector, and $G_k(M)$ is the effective BS antenna gain towards the $k$-th reflector. Then, $\Omega_A(K,\chi,M)$ is denoted as 
\begin{equation}
    \Omega_A(K,\chi,M)
    =
    \sum_{k=1}^{K}
    q_k(M)
    +
    2\chi
    \sum_{1\leq i<j\leq K}
    \sqrt{
    q_i(M)q_j(M)
    }.
    \label{eq:Omega_A_corrected}
\end{equation}
 
Similarly, we have
\begin{equation}
    g_T
    \sim
    \mathcal{CN}
    \left(
    0,
    \Omega_T(M)
    \right),
    \label{eq:gT_distribution_corrected}
\end{equation}
where
    $\Omega_T(M)
    =
    \sigma_T^2
    G_T(M)
    \mathcal{L}(d_{BT})
    \mathcal{L}(d_{TW})$, and $G_T(M)$ denotes the effective BS antenna gain towards the target.
    
Since $g_A$ and $g_T$ are independent zero-mean complex Gaussian random variables, their sum also follows a complex Gaussian distribution
\begin{equation}
    g_A+g_T
    \sim
    \mathcal{CN}
    \left(
    0,
    \Omega_A(K,\chi,M)+\Omega_T(M)
    \right).
    \label{eq:gA_plus_gT_distribution}
\end{equation}
Therefore, the random received powers under the two hypotheses follow exponential distributions
\begin{align}
    &Z_0
    =
    P_B|g_A|^2
    \sim
    \mathrm{Exp}
    \left(
    \bar{Z}_0
    \right),
     \\
    &Z_1
    =
    P_B|g_A+g_T|^2
    \sim
    \mathrm{Exp}
    \left(
    \bar{Z}_1
    \right),
\end{align}
where
    $\bar{Z}_0 = P_B
    \Omega_A(K,\chi,M) = \sigma_W^2
    \gamma_B
    \Omega_A(K,\chi,M)$,
    $\bar{Z}_1
    =
    P_B
    \left[
    \Omega_A(K,\chi,M)
    +
    \Omega_T(M)
    \right]
    =
    \sigma_W^2
    \gamma_B
    \left[
    \Omega_A(K,\chi,M)
    +
    \Omega_T(M)
    \right]$, and 
    $\gamma_B
    =
    \frac{P_B}{\sigma_W^2}$ is the BS transmit SNR.

For a given threshold $\tau$, Willie's false alarm probability is
    $P_{\mathrm{FA},W}(\tau)
    =
    \Pr
    \left(
    T_W>\tau
    \mid
    \mathcal{H}_0
    \right)$. Based on \eqref{eq:TW_H0_corrected}, we obtain
\begin{equation}
    P_{\mathrm{FA},W}(\tau)
    =
    \Pr
    \left(
    Z_0+\sigma_W^2>\tau
    \right).
    \label{eq:PFA_Z0}
\end{equation}
Since $Z_0$ is exponentially distributed with mean $\bar{Z}_0$, the false alarm probability is
\begin{equation}
    P_{\mathrm{FA},W}(\tau)
    =
    \begin{cases}
    1,
    & \tau<\sigma_W^2,\\[1mm]
    \exp
    \left(
    -
    \dfrac{\tau-\sigma_W^2}{\bar{Z}_0}
    \right),
    & \tau\geq\sigma_W^2.
    \end{cases}
    \label{eq:PFA_corrected}
\end{equation}

Similarly, Willie's missed detection probability is
    $P_{\mathrm{MD},W}(\tau)
    =
    \Pr
    \left(
    T_W\leq\tau
    \mid
    \mathcal{H}_1
    \right)$. Based on \eqref{eq:TW_H0_corrected}, we have
\begin{equation}
    P_{\mathrm{MD},W}(\tau)
    =
    \Pr
    \left(
    Z_1+\sigma_W^2\leq\tau
    \right).
    \label{eq:PMD_Z1}
\end{equation}
Since $Z_1$ is exponentially distributed with mean $\bar{Z}_1$, the missed detection probability is
\begin{equation}
    P_{\mathrm{MD},W}(\tau)
    =
    \begin{cases}
    0,
    & \tau<\sigma_W^2,\\[1mm]
    1-
    \exp
    \left(
    -
    \dfrac{\tau-\sigma_W^2}{\bar{Z}_1}
    \right),
    & \tau\geq\sigma_W^2.
    \end{cases}
    \label{eq:PMD_corrected}
\end{equation}

\subsection{Total Detection Error Probability Boundary}

The total detection error probability at Willie is
\begin{equation}
    P^*_E(\tau)
    =
    P_{\mathrm{FA},W}(\tau)
    +
    P_{\mathrm{MD},W}(\tau).
    \label{eq:PE_definition_corrected}
\end{equation}
Substituting \eqref{eq:PFA_corrected} and \eqref{eq:PMD_corrected} into \eqref{eq:PE_definition_corrected}, we obtain
For $\tau\geq \sigma_W^2$, define $t=\tau-\sigma_W^2$. Then, the total detection error probability is given by
\begin{equation}
    P_E(t)
    =
    1
    +
    \exp\left(-\frac{t}{\bar{Z}_0}\right)
    -
    \exp\left(-\frac{t}{\bar{Z}_1}\right),
    \quad t\geq 0,
    \label{eq:PE_t_before_optimization}
\end{equation}
and for $\tau<\sigma_W^2$, $P_E(\tau)=1$.

For $\tau\geq\sigma_W^2$, define
\begin{equation}
    t
    =
    \tau-\sigma_W^2.
    \label{33}
\end{equation}
Then, Willie's optimal probability of target detection  can be obtained by setting the derivative of $P^*_E(t)$ to zero $\frac{\partial P^*_E(t)}{\partial t}
    =
    -
    \frac{1}{\bar{Z}_0}
    \exp
    \left(
    -
    \frac{t}{\bar{Z}_0}
    \right)
    +
    \frac{1}{\bar{Z}_1}
    \exp
    \left(
    -
    \frac{t}{\bar{Z}_1}
    \right)
    =
    0$. Then, we have 
\begin{equation}
    P^*_E(t)
    =
    1
    +
    \exp
    \left(
    -
    \frac{t}{\bar{Z}_0}
    \right)
    -
    \exp
    \left(
    -
    \frac{t}{\bar{Z}_1}
    \right).
    \label{eq:PE_t_corrected}
\end{equation}
where    $t^\ast
    =
    \frac{
    \bar{Z}_0\bar{Z}_1
    }{
    \bar{Z}_1-\bar{Z}_0
    }
    \ln
    \left(
    \frac{\bar{Z}_1}{\bar{Z}_0}
    \right)$. According \eqref{33}, Willie's optimal energy threshold is
\begin{equation}
    \tau^\ast
    =
    \sigma_W^2
    +
    \frac{
    \bar{Z}_0\bar{Z}_1
    }{
    \bar{Z}_1-\bar{Z}_0
    }
    \ln
    \left(
    \frac{\bar{Z}_1}{\bar{Z}_0}
    \right).
    \label{eq:tau_star_corrected}
\end{equation}

Define the scattering separability ratio as
\begin{equation}
    r_W
    =
    \frac{\bar{Z}_1}{\bar{Z}_0}
    =
    1+
    \frac{
    \Omega_T(M)
    }{
    \Omega_A(K,\chi,M)
    }.
    \label{eq:rW_definition}
\end{equation}
Since $\bar{Z}_1>\bar{Z}_0$, we have $r_W>1$. Substituting $t^\ast$ into \eqref{eq:PE_t_corrected}, Willie's minimum total detection error probability is given by
\begin{equation}
    P^*_E
    =
    1
    +
    r_W^{-\frac{r_W}{r_W-1}}
    -
    r_W^{-\frac{1}{r_W-1}}.
    \label{eq:PE_star_corrected}
\end{equation}

\section{Performance Evaluation}

In this section, we present numerical experimental results to evaluate the effectiveness of the proposed reflector-assisted LPTD scheme and to validate the derived detection error boundary. Based on the typical setting in wireless sensing and reflector \cite{11358431,
3gpp_TR38901, famaey2026survey}, the BS, target, and Willie are located at $(0,0)$ m, $(50,0)$ m, and $(40,20)$ m, respectively. Passive reflectors are randomly distributed within a circular region of radius $5$ m centered at the target. The BS is equipped with a $16$-antenna ULA and steers its beam toward the target. The carrier frequency is set to $28$ GHz, the path-loss exponent is $\alpha=2.2$, and the reference path loss is given by $C_0=(\lambda/(4\pi d_0))^2$ with $d_0=1$ m. The noise floor at Willie is $-90$ dBm, and the BS transmit SNR is set to $10$ dB. The average reflector reflection power and partial coherence factor are set to $\rho^2=0.1$ and $\chi=0.2$, respectively. We consider three target reflection strengths, i.e., $\sigma_T^2={1,3,5}$, corresponding to small, medium, and large targets. The number of reflectors varies from $1$ to $29$. Each result is averaged over $100$ independent reflector layouts and $3\times 10^4$ Monte Carlo realizations.

\begin{figure}[tp]
    \centering
    \includegraphics[width=0.8\linewidth]{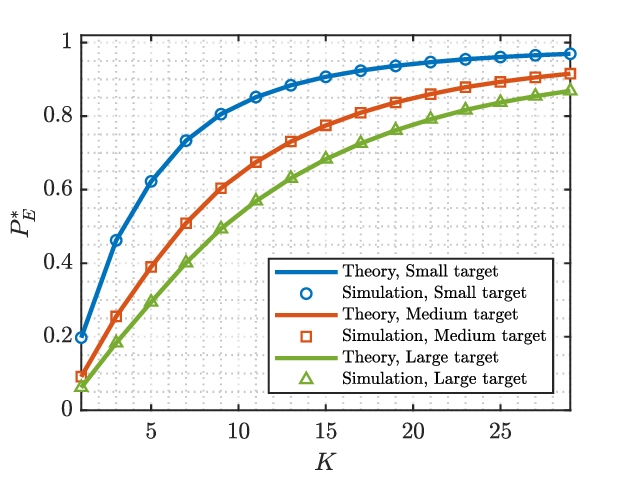}
    \caption{Impact of reflector Number on LPTD Performance.}
    \label{1}
   \vspace{-0.5cm}
\end{figure}
 
Fig.~\ref{1} compares the theoretical and simulation results of Willie's minimum total detection error probability under different numbers of reflectors. The close agreement between the simulation markers and theoretical curves validates the derived analytical results. Moreover, $P_E^\ast$ increases with $K$ because additional reflectors strengthen the surrounding scattering, reducing the statistical difference between $\mathcal{H}_0$ and $\mathcal{H}_1$ and thereby masking the target echo.
Smaller targets achieve higher $P_E^\ast$ since their weaker reflection produces a smaller $\Omega_T$ and is therefore easier to conceal. Although larger targets are more detectable, increasing the number of reflectors still significantly improves the LPTD performance, demonstrating the effectiveness of the proposed target covertness scheme.

\begin{figure}[tp]
    \centering
    \includegraphics[width=0.8\linewidth]{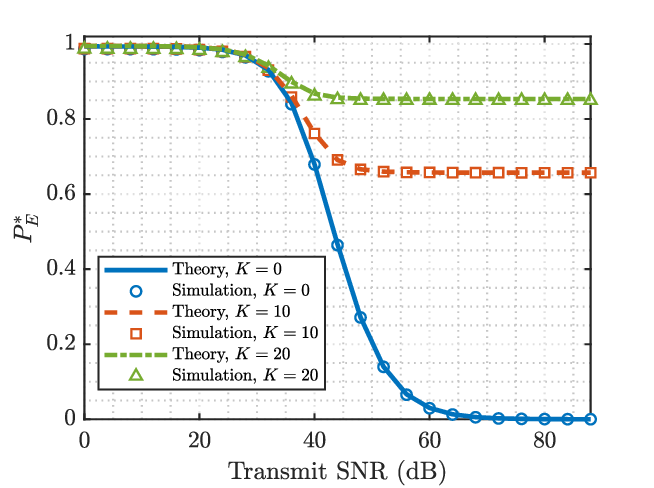}
    \caption{Impact of Transmit SNR on LPTD Performance.}
    \label{2}
   \vspace{-0.5cm}
\end{figure}

\begin{figure}[tp]
    \centering
    \includegraphics[width=0.8\linewidth]{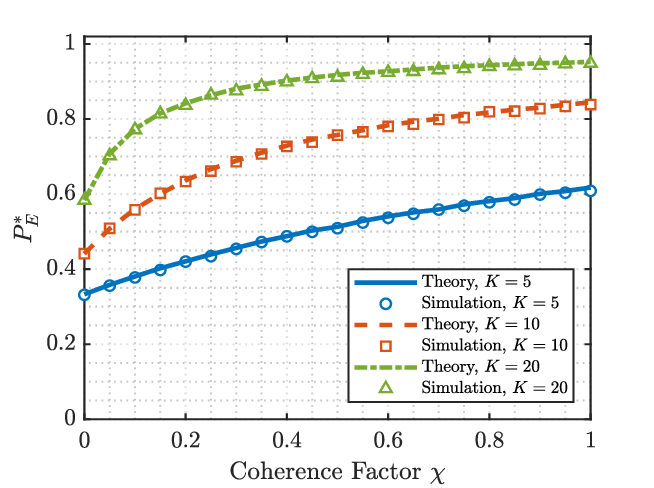}
    \caption{Impact of reflector Scattering Coherence on LPTD Performance.}
    \label{3}
   \vspace{-0.5cm}
\end{figure}

Fig.~\ref{2} illustrates the impact of the transmit SNR $\gamma_B$ on Willie's detection performance. Without reflector assistance, a finite-block condition \cite{11104928} is considered, where we find that increasing ($\gamma_B$) strengthens the target echo observed by Willie. As a result, the received energy distributions under ($\mathcal H_0$) and ($\mathcal H_1$) become increasingly separated, leading to a lower ($P_E^\ast$). This indicates that, in conventional wireless sensing without ambient scattering protection, a higher sensing transmit power directly benefits the passive target detector. In contrast, when reflectors are present, both the target-induced echo and the reflector-induced reflected signals are amplified by the BS transmit power. Hence, the dominant factor gradually changes from the absolute received power level to the relative scattering separability between the target echo and the reflector-induced ambient reflections. This explains why the reflector-assisted curves maintain much higher ($P_E^\ast$) over the considered SNR range.

Fig.~\ref{3} shows the effect of the coherence factor $\chi$ among reflector-reflected paths. It can be seen that $P^*_E$ increases as $\chi$ becomes larger, especially when more reflectors are deployed. This is because a larger $\chi$ enhances the coherent combination of the reflector-induced scattering components, thereby increasing the aggregate ambient scattering strength observed by Willie. As a result, the target-induced echo becomes less distinguishable from the surrounding reflected signals. 
The results indicate that not only the number of reflectors but also their scattering coherence plays an important role in determining the target covertness performance. In practice, this suggests that properly coordinating the reflection phases of reflectors can further enhance LPTD performance without increasing the BS transmit power.

Overall, the above results consistently demonstrate the effectiveness of the proposed reflector-assisted LPTD scheme. By increasing the number of cooperative reflectors and enhancing the aggregate ambient scattering strength, the received energy distributions under the target-absent and target-present hypotheses become highly overlapped at Willie. Under the considered parameter settings, ($P_E^\ast$) can be increased to about 0.94--0.96, which means that Willie has almost no reliable capability to distinguish whether the true target exists. This result confirms that reflectors can effectively create a scattering-dominated uncertainty region around the target, thereby enabling a LPTD without active jamming or additional dedicated deployment.

\section{Conclusion}

This paper proposed a novel reflector-assisted target covertness scheme to achieve an LPTD at the warden of the malicious target detector. Rather than relying on active jamming, we harnessed distributed passive reflectors to reshape the local scattering environment, thereby overwhelming the target's echo with controllable artificial uncertainty. Furthermore, we derived the closed-form boundary for the warden's target detection error, revealing the fundamental relationship between the covertness and key system parameters. The results showed that Willie’s minimum total detection error probability can be increased to around 0.94--0.96 with sufficient cooperation among reflectors, making the warden's target detection close to random guessing.

\section{Acknowledge}
This work was supported by the Smart Networks and Services Joint Undertaking under the EU’s Horizon Europe program (Ambient-6G, Grant No. 101192113), Business Finland (FISRE, Dnro 955/31/2026), and the Research Council of Finland (BANLOS, Decision No. 375519).


\bibliographystyle{IEEEtran}
\bibliography{reference}

@book{tse2005fundamentals,
  author    = {David Tse and Pramod Viswanath},
  title     = {Fundamentals of Wireless Communication},
  publisher = {Cambridge University Press},
  address   = {Cambridge, U.K.},
  year      = {2005}
}

@ARTICLE{9696209,
  author={Deng, Ruoqi and Di, Boya and Zhang, Hongliang and Tan, Yunhua and Song, Lingyang},
  journal={IEEE Transactions on Wireless Communications}, 
  title={Reconfigurable Holographic Surface-Enabled Multi-User Wireless Communications: Amplitude-Controlled Holographic Beamforming}, 
  year={Jan. 2022},
  volume={21},
  number={8},
  pages={6003-6017},
  doi={10.1109/TWC.2022.3144992}}

@article{zhang2025artificial,
  title={Artificial Noise Management for Securing Backscatter Communication Networks: Current Advances, Open Challenges, and Future Directions},
  author={Zhang, Yifan and Yan, Zheng and Wang, Jie and Yang, Yishan and Li, Zhao and J{\"a}ntti, Riku},
  journal={IEEE Network},
  year={Early access, May 2025},
  publisher={IEEE}
}

@article{zhang2024covert,
  title={Covert backscatter communication in the presence of multi-antenna eavesdropper},
  author={Zhang, Yifan and Di Candia, Roberto and Yi{\u{g}}itler, H{\"u}seyin and J{\"a}ntti, Riku and Yan, Zheng},
  journal={IEEE Communications Letters},
  volume={28},
  number={8},
  pages={1770--1774},
  year={Jun. 2024},
  publisher={IEEE}
}

@ARTICLE{10109228,
  author={Li, Xingwang and Jiang, Junjie and Wang, Hao and Han, Congzheng and Chen, Gaojie and Du, Jianhe and Hu, Chunqiang and Mumtaz, Shahid},
  journal={IEEE Transactions on Cognitive Communications and Networking}, 
  title={Physical Layer Security for Wireless-Powered Ambient Backscatter Cooperative Communication Networks}, 
  year={Apr. 2023},
  volume={9},
  number={4},
  pages={927-939},
  doi={10.1109/TCCN.2023.3270425}}

@ARTICLE{11202391,
  author={Han, Kawon and Meng, Kaitao and Masouros, Christos},
  journal={IEEE Transactions on Wireless Communications}, 
  title={Sensing-Secure {ISAC}: Ambiguity Function Engineering for Impairing Unauthorized Sensing}, 
  year={Oct. 2025},
  volume={},
  number={},
  pages={1-1},
  doi={10.1109/Transactions on Wireless Communications.2025.3618121}}

@ARTICLE{10587082,
  author={Zou, Jiaqi and Masouros, Christos and Liu, Fan and Sun, Songlin},
  journal={IEEE Transactions on Vehicular Technology}, 
  title={Securing the Sensing Functionality in {ISAC} Networks: An Artificial Noise Design}, 
  year={Jul. 2024},
  volume={73},
  number={11},
  pages={17800-17805},
  doi={10.1109/TVT.2024.3422036}}

@ARTICLE{11155198,
  author={Zhang, Zhentian and Wong, Kai-Kit and Dang, Jian and Zhang, Zaichen and Chae, Chan-Byoung},
  journal={IEEE Journal on Selected Areas in Communications}, 
  title={On Fundamental Limits for Fluid Antenna-assisted Integrated Sensing and Communications for Unsourced Random Access}, 
  year={Sep. 2025},
  volume={},
  number={},
  pages={1-1},
  doi={10.1109/JSAC.2025.3608113}}

@ARTICLE{Zhang2026AmbShield,
  author={Zhang, Yifan and Yang, Yishan and Kaveh, Masoud and Jäntti, Riku and Yan, Zheng and Niyato, Dusit and Han, Zhu},
  journal={IEEE Transactions on Information Forensics and Security}, 
  title={AmbShield: Enhancing Physical Layer Security With Ambient Backscatter Devices Against Eavesdroppers}, 
  year={Apr. 2026},
  volume={21},
  number={},
  pages={4547-4562},
  doi={10.1109/TIFS.2026.3688905}}

@ARTICLE{zhang2025backscatter,
  author={Zhang, Yifan and Bai, Yu and Zeng, Shuhao and Jäntti, Riku and Yan, Zheng and Masouros, Christos and Han, Zhu},
  journal={IEEE Transactions on Wireless Communications}, 
  title={Backscatter Device-Aided Integrated Sensing and Communication: A Pareto Optimization Framework}, 
  year={Jun. 2026},
  volume={25},
  number={},
  pages={17958-17974},
  doi={10.1109/Transactions on Wireless Communications.2026.3698666}}

@ARTICLE{11071870,
  author={Zhang, Yifan and Xie, Boxuan and Yang, Yishan and Yan, Zheng and Jäntti, Riku and Han, Zhu},
  journal={IEEE Transactions on Information Forensics and Security}, 
  title={AuthScatter: Accurate, Robust, and Scalable Mutual Authentication in Physical Layer for Backscatter Communications}, 
  year={Jul. 2025},
  volume={20},
  number={},
  pages={6937-6952},
  doi={10.1109/TIFS.2025.3585453}}

@ARTICLE{11104928,
  author={Zhang, Zhentian and Wong, Kai-Kit and Dang, Jian and Zhang, Zaichen and Masouros, Christos and Chae, Chan-Byoung},
  journal={IEEE Wireless Communications Letters}, 
  title={On Fundamental Limits of Slow-Fluid Antenna Multiple Access for Unsourced Random Access}, 
  year={Jul. 2025},
  volume={14},
  number={11},
  pages={3455-3459},
  doi={10.1109/LWC.2025.3594112}}

@ARTICLE{10436390,
  author={Zhang, Zheng and Liu, Yuanwei and Wang, Zhaolin and Mu, Xidong and Chen, Jian},
  journal={IEEE Transactions on Vehicular Technology}, 
  title={Physical Layer Security in Near-Field Communications}, 
  year={Feb. 2024},
  volume={73},
  number={7},
  pages={10761-10766},
  doi={10.1109/TVT.2024.3366115}}

@article{famaey2026survey,
  title={A survey on ambient {IoT}: Use cases, standardization, enabling technologies, and research directions},
  author={Famaey, Jeroen and L{\'o}pez, Onel LA and P{\'e}rez, Rub{\'e}n Altuna and Callebaut, Gilles and Strycker, Lieven De and Katranaras, Efstathios and Liao, Jingyi and Singh, Ritesh Kumar and Ashraf, Mateen and Azarbahram, Amirhossein and others},
  year={Feb. 2026},
  journal={TechRxiv}
}

@ARTICLE{11122381,
  author={Bai, Jiale and Wang, Hui-Ming and Jin, Liang},
  journal={IEEE Transactions on Wireless Communications}, 
  title={Dynamic Agile Reconfigurable Intelligent Surface Antenna ({DARISA}) {MIMO}: {DoF} Analysis and Effective DoF Optimization}, 
  year={Aug. 2025},
  volume={25},
  number={},
  pages={2197-2212},
  doi={10.1109/TWC.2025.3595260}}

@ARTICLE{10497104,
  author={Zhang, Yuchen and Ni, Wanli and Wang, Jianquan and Tang, Wanbin and Jia, Min and Eldar, Yonina C. and Niyato, Dusit},
  journal={IEEE Transactions on Communications}, 
  title={Robust Transceiver Design for Covert Integrated Sensing and Communications With Imperfect CSI}, 
  year={Apr. 2025},
  volume={73},
  number={9},
  pages={8016-8031},
  doi={10.1109/TCOMM.2024.3387869}}

@ARTICLE{10906391,
  author={Yuchen, Zhang and Sa, Xiao and Jianquan, Wang and Boyu, Ning and Xiaojun, Yuan and Wanbin, Tang},
  journal={China Communications}, 
  title={Covert rate characterization under multi-antenna detection: A study on warden's CSI availability and noise uncertainty}, 
  year={Apr. 2025},
  volume={22},
  number={2},
  pages={143-159},
  doi={10.23919/JCC.ja.2023-0004}}

@ARTICLE{11318877,
  author={Li, Zan and Hui, Pei and Guan, Lei and Li, Chenxi},
  journal={IEEE Transactions on Communications}, 
  title={A Frequency Hopping Scheme for Covert Communication With Noise Uncertainty}, 
  year={Dec. 2025},
  volume={74},
  number={},
  pages={3132-3146},
  doi={10.1109/TCOMM.2025.3649718}}

@ARTICLE{11365966,
  author={Wang, Chao and Li, Zan and Zhou, Xiangnan and Jin, Liang},
  journal={IEEE Journal of Selected Topics in Signal Processing}, 
  title={Intelligent Cooperative Covert ISAC via Fluid Antennas: A GAN-Optimized Framework}, 
  year={Jan. 2026},
  volume={20},
  number={3},
  pages={262-276},
  doi={10.1109/JSTSP.2026.3658184}}

@book{poor2013introduction,
  title={An introduction to signal detection and estimation},
  author={Poor, H Vincent},
  year={2013},
  publisher={Springer Science \& Business Media}
}

@ARTICLE{7964713,
  author={Sobers, Tamara V. and Bash, Boulat A. and Guha, Saikat and Towsley, Don and Goeckel, Dennis},
  journal={IEEE Transactions on Wireless Communications}, 
  title={Covert Communication in the Presence of an Uninformed Jammer}, 
  year={Jun. 2017},
  volume={16},
  number={9},
  pages={6193-6206},
  doi={10.1109/TWC.2017.2720736}}

@ARTICLE{luoniya,
  author={Niya Luo, Yishan Yang, Zheng Yan and Yifan Zhang},
  journal={IEEE Communications Surveys \& Tutorials}, 
  title={Towards Secure Reflection: A Survey on Physical Layer Security in Backscatter Communications}, 
  year={Jun. 2026},
  pages={1-34}}

@techreport{3gpp_TR38901,
  author      = {{3rd Generation Partnership Project (3GPP)}},
  title       = {Study on channel model for frequencies from 0.5 to 100 {GHz}},
  institution = {3GPP},
  type        = {Technical Report (TR)},
  number      = {38.901},
  year        = {2020},
  note        = {Release 16}
}

@ARTICLE{11667367,
  author={Famaey, Jeroen and López, Onel L. A. and Pérez, Rubén Altuna and Callebaut, Gilles and De Strycker, Lieven and Katranaras, Efstathios and Liao, Jingyi and Singh, Ritesh Kumar and Ashraf, Mateen and Azarbahram, Amirhossein and Belogaev, Andrey and Cappelle, Jona and Chassaigne, Alexander and Moura, Henrique Duarte and Deutschmann, Benjamin and Filippou, Miltiadis C. and Jahanbazi, Shahab and Jäntti, Riku and Kaveh, Masoud and Leenders, Guus and Muehlmann, Ulrich and Nasser, Samer and Puluckul, Priyesh Pappinisseri and Poehl, Daniel and Raghuwanshi, Prasoon and Ramaut, Lyssa and Rosabal, Osmel M. and Ruiz-Guirola, David E. and Singh, Bikramjit and Xie, Boxuan and Zhang, Yifan},
  journal={IEEE Communications Surveys \& Tutorials}, 
  title={A Survey on Ambient IoT: Use Cases, Standardization, Enabling Technologies, and Research Directions}, 
  year={Aug. 2026},
  volume={},
  number={},
  pages={1-1},
  doi={10.1109/COMST.2026.3727751}}

@ARTICLE{10955353,
  author={Gan, Lin and Wu, Zehao and Li, Jianbing},
  journal={IEEE Transactions on Aerospace and Electronic Systems}, 
  title={A Passive ISAR Jamming Method Based on Time-Modulated Corner Reflectors Under Asynchronous Control}, 
  year={Apr. 2025},
  volume={61},
  number={5},
  pages={11103-11115},
  doi={10.1109/TAES.2025.3556413}}

@ARTICLE{11358431,
  author={Zhang, Yifan and Dang, Yongchao and Kaveh, Masoud and Yan, Zheng and Jäntti, Riku and Han, Zhu},
  journal={IEEE Transactions on Information Forensics and Security}, 
  title={Physical Layer Challenge-Response Authentication Between Ambient Backscatter Devices}, 
  year={Jan. 2026},
  volume={21},
  number={},
  pages={1512-1527},
  doi={10.1109/TIFS.2026.3655191}}

@article{zhang2026ambsentry,
  title={AmbSentry: Mitigating Sensing Eavesdropping in ISAC Systems by Harnessing Ambient IoT Devices},
  author={Zhang, Yifan and Bai, Yu and Jantti, Riku and Han, Zhu and Masouros, Christos},
  journal={arXiv preprint arXiv:2608.11799},
  year={2026}
}

\end{document}